\documentclass[aps,pre,showpacs]{revtex4}

\usepackage{graphicx}
\usepackage{amsmath}

\newcommand{\be}{\begin{equation}}
\newcommand{\ee}{\end{equation}}
\newcommand{\bea}{\begin{eqnarray}}
\newcommand{\eea}{\end{eqnarray}}

\newcommand \lan {\langle} 
\newcommand \ran {\rangle} 

\newcommand{\br}{\mathbf{r}}
\newcommand{\bR}{\mathbf{R}}

\newcommand{\bt}{\mathbf{t}}

\newcommand{\bK}{\mathbf{K}}

\newcommand{\md}{\mathrm{d}}
\newcommand{\kB}{k_{\rm B}}

\def\eq#1{Eq.~(\ref{#1})}
\def\eqs#1#2{Eqs.~(\ref{#1},\ref{#2})}

\def\w3j#1#2#3#4#5#6{\left( 
\begin{array}{ccc}
{#1} & {#2} & {#3} \\
{#4} & {#5} & {#6}
\end{array}
\right)}
\def\lmax{\ell_{\rm max}}

\begin{document}

\title{Mesoscopic models for DNA stretching under force:\\ new results and comparison to experiments}

\author{Manoel Manghi}
\email{manghi@irsamc.ups-tlse.fr}
\author{Nicolas Destainville}
\author{John Palmeri}

\affiliation{                    
(1) Universit\'e de Toulouse; UPS; Laboratoire de Physique Th\'eorique (IRSAMC); F-31062 Toulouse, France, EU\\
(2) CNRS; LPT (IRSAMC); F-31062 Toulouse, France, EU
}

\begin{abstract}
Single molecule experiments on double stranded B-DNA stretching have revealed one or two structural transitions, when increasing the external force. They are characterized by a sudden increase of DNA contour length and a decrease of the bending rigidity. The nature and the critical forces of these transitions depend on DNA base sequence, loading rate, salt conditions and temperature.  It has been proposed that the first transition, at forces of 60--80~pN, is a transition from B to S-DNA, viewed as a stretched duplex DNA, while the second one, at  stronger forces, is a strand peeling resulting in single stranded DNAs (ssDNA), similar to thermal denaturation. But due to experimental conditions these two transitions can overlap, for instance for poly(dA-dT). In an attempt to propose a coherent picture compatible with this variety of experimental observations, we derive analytical formula using a coupled discrete worm like chain-Ising model. Our model takes into account bending rigidity, discreteness of the chain, linear and non-linear (for ssDNA) bond stretching. In the limit of zero force, this model simplifies into a coupled model already developed by us for studying thermal DNA melting, establishing a connexion with previous fitting parameter values for denaturation profiles. Our results are summarized as follows: (i) ssDNA is fitted, using an analytical formula, over a nanoNewton range with only three free parameters, the contour length, the bending modulus and the monomer size; (ii) a surprisingly good fit on this force range is possible only by choosing a monomer size of 0.2~nm, almost 4 times smaller than the ssDNA nucleobase length; (iii) mesoscopic models are not able to fit B to ssDNA (or S to ss) transitions; (iv) an analytical formula for fitting B to S transitions is derived in the strong force approximation and for long DNAs, which is in excellent agreement with exact transfer matrix calculations; (v) this formula fits perfectly well poly(dG-dC) and $\lambda$-DNA force-extension curves with consistent parameter values; (vi) a coherent picture, where S to ssDNA transitions are much more sensitive to base-pair sequence than the B to S one, emerges. This relatively simple model might allow one to further study quantitatively the influence of salt concentration and base-pairing interactions on DNA force-induced transitions.
\end{abstract}

\maketitle

\section{Introduction}
\label{intro}

In the recent decades, many experimental developments have been devoted to the manipulation and analysis of single molecules such as nucleic acids and proteins, proving to be invaluable tools to understand their statistical and mechanical properties~\cite{Neuman08}. They include atomic force microscopy (AFM)~\cite{rief}, fluorescence microscopy~\cite{maier}, tethered particle motion~\cite{shafer}, as well as magnetic and optical tweezers~\cite{Cluzel96,Smith96,Bustamante03,Woodside08}. Understanding the mechanical properties of nucleic acids at the nanometric scale is crucial because they play a role in both their biological functions and their packaging in a variety of circumstances, such as interaction with other macromolecules (e.g. histones or ribosomes), DNA cyclization, DNA looping in some genetic regulatory processes, or nucleic acid packing in viruses. Furthermore, single-molecule measurements on nucleic acids give the possibility to investigate their interactions with partner proteins~\cite{Neuman08,Woodside08}. 

The present work primarily focuses on the response of nucleic acids to an external force applied without torsional constraint, e.g., by optical tweezers or AFM. Such experiments on double-stranded DNA (dsDNA)  at room temperature show a sharp, few picoNewtons wide, cooperative overstretching ``transition''~\footnote{This transition is not a true, thermodynamical transition \textit{stricto sensu}, but this terminology is widely used in the field.} at a given ``critical' force of around 60--80~pN, accompanied by a sudden 70~\% increase of the contour length~\cite{Cluzel96,Smith96}. 

By applying force up to 800~pN, Rief \textit{et al.}~\cite{rief,clausen} measured a second transition at stronger forces, which is hysteretic and consistent with a peeling of one strand from its complementary strand. They show moreover that the critical force of this second transition depends dramatically on the DNA sequence, from 150~pN for $\lambda$-phage DNA to 320~pN for poly(dG-dC). They did not measure any 2d transition for poly(dA-dT). 

The nature of the first transition at  60--80~pN remains controversial because it is not definitely established whether it is a transition from the native B-form to a new form of unstacked DNA remaining in a duplex form (the so-called ``S'' form for Stretched), or double-strand separation leading to two single stranded DNA (ssDNA) similar to the thermal denaturation. This controversy~\cite{Rouzina01} is reviewed in detail, e.g., in the Refs.~\cite{williams,vanMameren09,Netz10,williams2}. 
Some authors even argue that the nature of the transition depends on the loading rate, the slowest rates enabling equilibration and thus denaturation under force~\cite{Prevost09,Albrecht08}. In a very recent work~\cite{fu1,fu2,zhang}, two different transitions, both occurring at 60--80~pN for $\lambda$-phage DNA, have been revealed experimentally, one is a hysteretic strand peeling whereas the other is a nonhysteretic transition that leads to S-DNA. The selection between this two transition depends on the DNA sequence and the salt concentration. Such study may thus reconcile the previous ones, once a careful comparison of both the experimental salt concentrations and the sequence of the studied DNAs will be done. 

Previous arguments for a B to S-DNA transition was that the part of the force-extension curve beyond the transition does not fit with a Worm-like Chain (WLC) model with the bending modulus and the monomer size of ssDNA~\cite{Cluzel96,Smith96,cocco}. However, the simpler stretching experiment on single ssDNA is already not easily fitted by the WLC model. Several explanations have been put forward,  such as electrostatic interactions~\cite{dessinges}, discreteness of the chain~\cite{livadaru}, and non-linear bond stretching~\cite{hugel}. Using a variational continuous WLC model with linear stretching, Storm and Nelson~\cite{storm1,storm2}  were able to fit ssDNA for forces smaller than 0.1~nN but with an extremely low value for the monomer size between 0.1 and 0.25~nm. 

Generally, the challenge is to propose a consensual mechanism compatible with this variety of experimental observations. For instance, no adequate analytical formula exists for fitting dsDNA force-extension curves including the transitions. Since the experimental force-extension curves are measured for forces that range from 0 to 800~nN, the model should scan both the low force regime where bending and entropic effects are important and the strong force regime where bond extension is non-negligible. Existing theoretical works use thermodynamical approaches~\cite{Rouzina01}, interpolations~\cite{cizeau,ahsan} between two (semi-)flexible chains using the Marko-Siggia interpolation formula~\cite{Marko95}, generalized Poland-Scheraga models with external force~\cite{hanke1,rudnick}, two-state continuous WLC models treated variationally~\cite{storm1,storm2}, or three-state models~\cite{cocco,Netz10}.

In the present paper, we first focus on ssDNA force-extension curves and provide an accurate analytical formula for fitting the stretching curves up to 1~nN (Section~\ref{sec_ssDNA}). This formula includes bending rigidity (which plays a central role in the low force regime), discreteness of the chain, and non-linear stretching already studied by H\"ugel \textit{et al.}~\cite{hugel} (which has been shown to be very important for forces stronger than a few hundreds of pN). We then show that describing a hypothetic B to ssDNA transition with models that do not consider explicitly the increase of the number of sub-nucleobase degrees of freedom scanned in the high force regime is out of reach.

Next, we use, in Section~\ref{sec_BtoSDNA}, a mesoscopic approach to model the B to S transition. This is a coupled Ising/Heisenberg model which is an extension of the mesoscopic model that we had introduced in 2007 for the description of temperature-induced melting of dsDNA~\cite{prl07,pre08}. This model is first solved exactly in Section~\ref{sec_BtoSDNA_exact} using pseudo-analytic transfer matrix calculations, following the same formalism as Rahi \textit{et al.}~\cite{Rahi08}. We then propose a simple analytical formula in the strong force approximation (Section~\ref{sec_BtoSDNA_sfa}), which is in excellent agreement with the exact results.

Finally, we compare our formulas (the fitting procedure are summarized in the appendix) to experimental force-extension curves in Section~\ref{sec_BtoSDNA_comp} and show surprisingly good agreement with several data sets for poly(dG-dC) and $\lambda$-DNAs. Since our model is on the same footing as the one describing DNA thermal denaturation, we propose an analytical formula for the ``coexistence'' line in the temperature-force diagram. Our final remarks are given in the Conclusion.

\section{Single strand DNA stretching}
\label{sec_ssDNA}

Before considering the transitions observed in dsDNA stretching experiments, we focus on ssDNA which is a good candidate for modeling a semiflexible chain under external load from 0 to 1~nN. At very large forces, the discrete nature of the polymer is probed, and it has been shown that force-extension curves are satisfactorily modeled by a freely jointed chain (FJC) model in the strong force regime~\cite{livadaru,rief,hugel,lipowsky,Rosa}. We thus focus on the discrete worm-like chain model (WLC), where the chain made of $N$ monomers of size $a$ is described by the effective Hamiltonian
\be
\mathcal{H}_{\rm DWLC}=\sum_{i=1}^{N-1}\kappa_b (1-\mathbf{t}_i\cdot\mathbf{t}_{i+1}) - a f \bt_i \cdot\hat z
\label{HDWLC}
\ee
where $\bt_i$ is the normalized orientation of monomer $i$, and $\kappa_b$ is the bending elastic modulus. The applied force is taken to be along $z$, $\mathbf{f} = f \hat z$.  The partition function is given by
\be
\mathcal{Z}=\prod_{i=1}^{N-1} a^2 \int \md\bt_i\; e^{-\beta \mathcal{H}_{\rm DWLC}}
\ee
The factor $a^2$ is the entropic contribution of the integration factor in phase space $\int \md (\br_{i+1}-\br_i) \delta(|\br_{i+1}-\br_i|-a) (\ldots)=  a^2 \int\md \bt(\ldots)$. 
The extension of the polymer along the direction of the force $\mathbf{f}$ is given by
\be
z\equiv \lan \hat z\cdot [\br_N-\br_1]\ran = \frac{\partial \ln\mathcal{Z}}{\partial \beta f}
\label{def_ext}
\ee
The mean-squared end-to-end distance at zero force in the limit $N\to\infty$ reads~\cite{pre08}
\be
\lan \bR^2\ran= a^2N\frac{1+u(\kappa)}{1-u(\kappa)}
\ee
where $u(x)=\coth(x)-1/x$ is the Langevin function and $\kappa=\beta\kappa_b$ is the adimensional bending modulus. The discrete persistence length is thus $\ell_p=-a/\ln u(\kappa)$. Since we consider a semi-flexible chain, we assume $\kappa\gtrsim1$. 

At small forces, such that the chain is only slightly deformed in the $z$ direction, the chain can be viewed as a linear string of Pincus blobs~\cite{pincus} of size $\xi=1/(\beta f)$ with $L>\xi>\ell_p>a$, where $L=aN$ is the contour length of the polymer. Hence in this regime where $\beta a\kappa f<1$, the linear response theory is valid and yields for the extension 
\be
z\simeq \kB T \left.\frac{\partial^2 \ln\mathcal{Z}}{\partial f^2}\right|_{f=0} f= \frac{\lan \bR^2\ran}{3\kB T} f \quad\mathrm{for}\quad f< \bar f\equiv\frac{\kB T}{a\kappa}
\label{small_forces}
\ee
Note that this relation is modified in good solvent according to the scaling law $z\simeq L(f/\bar f)^{2/3}$. In the following we neglect the eventual polymer swelling for DNA in water due to van der Waals~\cite{pincus} or electrostatic interactions~\cite{joanny}. Moreover the electrostatic contribution to the persistence length which might be important for flexible chains such as ssDNA is taken implicitly into account by choosing $\kappa$ as a fitting parameter~\cite{mano,netz1}.

At very large forces, $f\gg 4\kappa_b/a$, we see from \eq{HDWLC} that the bending energy can be neglected in the Hamiltonian and the freely jointed chain model is valid. The force probes the discrete nature of the chain since the Pincus blob size is much smaller than the monomer size a, $\xi\ll \kB T/(4\kappa_b a)$. One thus finds the classical Langevin result~\cite{Flory89}
\be
z\simeq L\; u(a\beta f) \simeq L \left(1-\frac1{a\beta f}\right)  \quad\mathrm{for}\quad  f\gg\frac{\kB T}{a}
\label{fjc}
\ee

For intermediate forces, $(\kB T)^2/(a\kappa_b)< f<4\kappa_b/a$, the two terms in \eq{HDWLC} should be taken into account, although the bending energy is negligible in the $z$ direction. Following Marko and Siggia~\cite{Marko95}, we use the approximation of large forces, which states that $\bt_i$ is mostly along $\hat z$ ($|t_{iz}|\gg|t_{ix}|,|t_{iy}|$) and thus $|t_{iz}|=\sqrt{1-t_{ix}^2-t_{iy}^2}\simeq1-(t_{ix}^2+t_{iy}^2)/2$. By noting that $\kappa(1-\mathbf{t}_i\cdot\mathbf{t}_{i+1})=\frac12\kappa(\mathbf{t}_i-\mathbf{t}_{i+1})^2$, the partition function can be rewritten as $\mathcal{Z}=\prod_{i=1}^{N-1}\int\md \bt_i\md \bt_{i+1} \hat P(\bt_i,\bt_{i+1})$ where $\hat P(\bt_i,\bt_{i+1})\equiv a^2 e^F K(t_{x,i},t_{x,i+1})K(t_{y,i},t_{y,i+1})$ is the transfer operator to be diagonalized.
The eigenvalue equation is
\be
a^2e^F\int_{-\infty}^{\infty}\md t_x' K(t_x,t_x')\, \phi_x(t_x') \int_{-\infty}^{\infty}\md t_y'K(t_y,t_y')\, \phi_y(t_y') =\Lambda \phi_x(t_x) \phi_y(t_y)
\label{Kern}
\ee
where we define the adimensional force $F=a\beta f$ and
\be
K(t,t')= \exp\left[-\frac{\kappa}2(t-t')^2-\frac{F}4(t^2+ t'^2)\right] 
\label{TOH}
\ee
This transfer kernel problem has first been solved by Fixman and Kovac~\cite{fixman} using a mode decomposition for a discrete worm like chain with extensible bonds (without applied force), and then extended to the force case in Ref.~\cite{Rosa,lipowsky}. For sake of clarity and as an introduction for the more complex case developed in the next section, where the mode decomposition is not applicable, we re-derive the calculation in the following by using a different approach, i.e. solving \eq{Kern} directly in real space.

We define $\phi_1(t)$ and $\phi_2(t')$ such that
\be
\int_{-\infty}^{\infty}\md t'  \exp\left[-\frac{\kappa}2(t-t')^2-\frac{F_1}4t^2-\frac{F_2}4 t'^2\right] \, \phi_2(t') = \lambda  \phi_1(t)
\label{kernel1}
\ee
If $\phi_1(t)= \exp(-\alpha_1t^2/2)$ and $\phi_2(t')= \exp(-\alpha_2t'^2/2)$, we find 
\be
\alpha_1=\kappa+\frac{F_1}2-\frac{\kappa^2}{\kappa+\frac{F_2}2+\alpha_2}\qquad\mathrm{and}\qquad \lambda=\sqrt{\frac{2\pi}{\kappa+\frac{F_2}2+\alpha_2}}
\label{result1}
\ee
If $F_1=F_2$, then $\phi_1=\phi_2$ is an eigenfunction for $\alpha_1=\alpha_2=\alpha=\sqrt{(F/2)^2+\kappa F}$ and $\Lambda=\lambda^2$.
The partition function in the limit $N\to\infty$ is thus
\be
\mathcal Z\simeq e^{NF} \left(\frac{2\pi a^2}{\kappa+\frac{F}2+\sqrt{\left(\frac F2\right)^2+\kappa F}}\right)^N
\ee
and the extension defined in \eq{def_ext} is
\be
 z  = L \left(1- \frac1{\sqrt{F^2+4\kappa F}}\right)
\label{ext_homo}
\ee
Again this formula is only valid for large forces, i.e. for $F\gg 1/\kappa$. Note that \eq{ext_homo} encompasses the very strong force regime $F\gg 4\kappa$ where the result of \eq{fjc} is recovered. 

An interpolation formula for the whole range of forces can be obtained following Marko and Siggia~\cite{Marko95} by inverting \eq{ext_homo}, subtracting the first two terms of the small $z$ expansion of $F(z/L)$, and then adding the zero force limit \eq{small_forces}. This yields the discrete Marko-Siggia interpolation~\cite{lipowsky}:
\be
F_{\rm DMS}\equiv \frac{a f_{\rm DMS} }{\kB T}= \frac{z}{L} \left(3\frac{1-u(\kappa)}{1+u(\kappa)}-\frac1{\sqrt{1+4\kappa^2}}\right)+\sqrt{\frac1{(1-z/L)^2}+4\kappa^2}-\sqrt{1+4\kappa^2}
\label{interp1}
\ee

Experimentally, one observes that, at very large forces, $f>k_BT/a$, the DNA starts to stretch elastically and the FJC result, \eq{fjc}, must be corrected so that $z >L$. Different models try to take this into account by using a linear correction~\cite{Marko95}, or using an extensible DWLC model~\cite{lipowsky}. However, it has been shown recently by H\"ugel \textit{et al.} that, beyond elastic stretching, non-linear terms are necessary to fit force-extension curves of peptides or polyvinyl-amine as well as ssDNA~\cite{hugel}. A consistent way is to modify \eq{ext_homo} according to
\be
z= L\left[1+U_{\rm nl}(f)\right]\left( 1- \frac1{\sqrt{F^2+4\kappa F}}\right)
\label{ext_stretch}
\ee
where 
\be
U_{\rm nl}(f) = 1.172777\; f  - 3.731836\;f^2 +  4.118249\; f^3 \qquad (f \;\mbox{in units of 10~nN})
\label{UNL}
\ee
is extracted from Ref.~\cite{hugel} [by inverting their Eq.~(2)] and is the result of \textit{ab-initio} quantum calculations for ssDNA. Corrections are on the order of $\approx 1\%$ for $f\simeq100$~pN, of $\approx 10\%$ for $\simeq1$~nN.

To reconcile the low force regime where \eq{interp1} is correct and the large force regime, \eq{ext_stretch}, where non-linear elasticity is non-negligible, we fit the experimental data for ssDNA using the interpolation \eq{interp1} with $z$ replaced by $z(1+U_{\rm nl}(f))$ in the rhs. It yields:
\be
\frac{a f}{\kB T}= \frac{z}{L} (1+U_{\rm nl}(f)) \left(3\frac{1-u(\kappa)}{1+u(\kappa)}-\frac1{\sqrt{1+4\kappa^2}}\right)+\sqrt{\frac1{[1-z(1+U_{\rm nl}(f))/L]^2}+4\kappa^2}-\sqrt{1+4\kappa^2}
\label{interp}
\ee
This non-linear equation is easily plotted using a parametric plot if we replace $U_{\rm nl}(f)$ by $U_{\rm nl}(f_{\rm DMS})$, where $f_{\rm DMS}$ is defined in \eq{interp1}. We have checked that this approximation is extremely good for forces in the nanoNewton range. We use \eq{interp} to fit experimental data with only three fitting parameters, namely the polymer contour length $L$, the adimensional bending modulus $\kappa$ and the monomer size $a$.
\begin{figure}[t]
\includegraphics[width=0.48\linewidth]{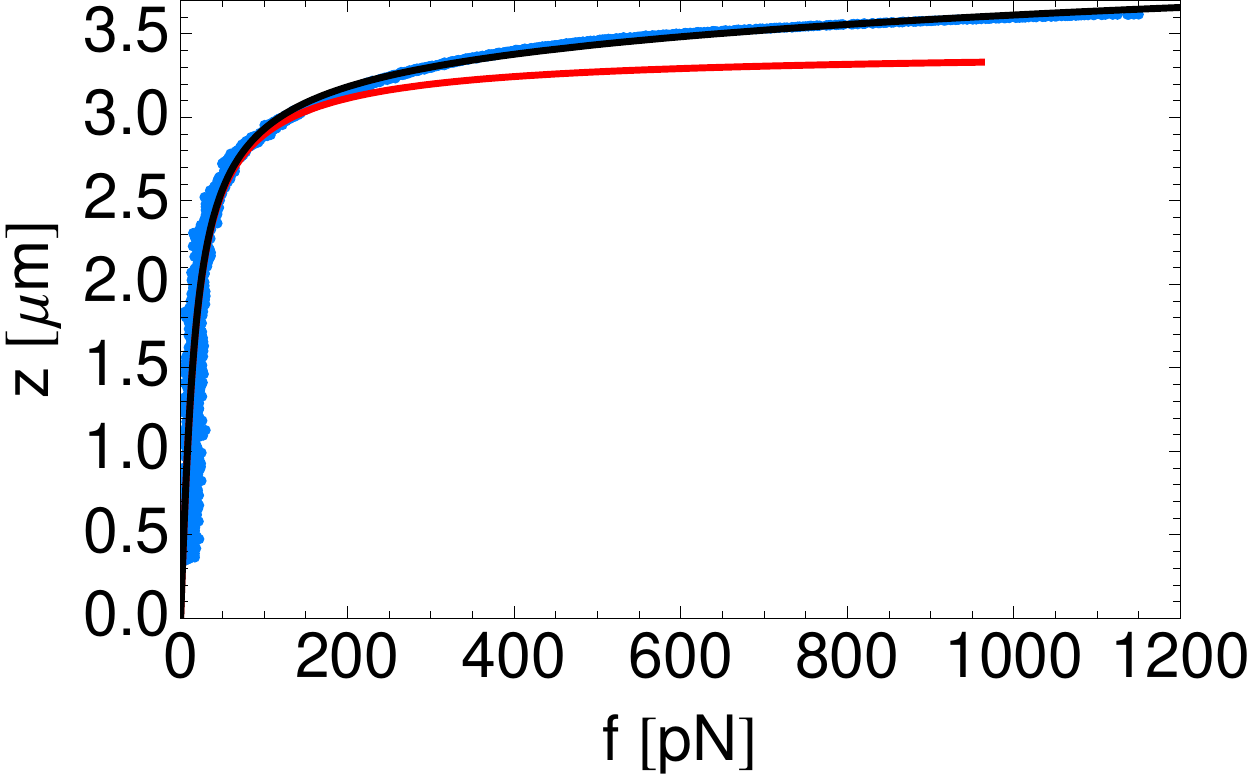}\hfill\includegraphics[width=0.47\linewidth]{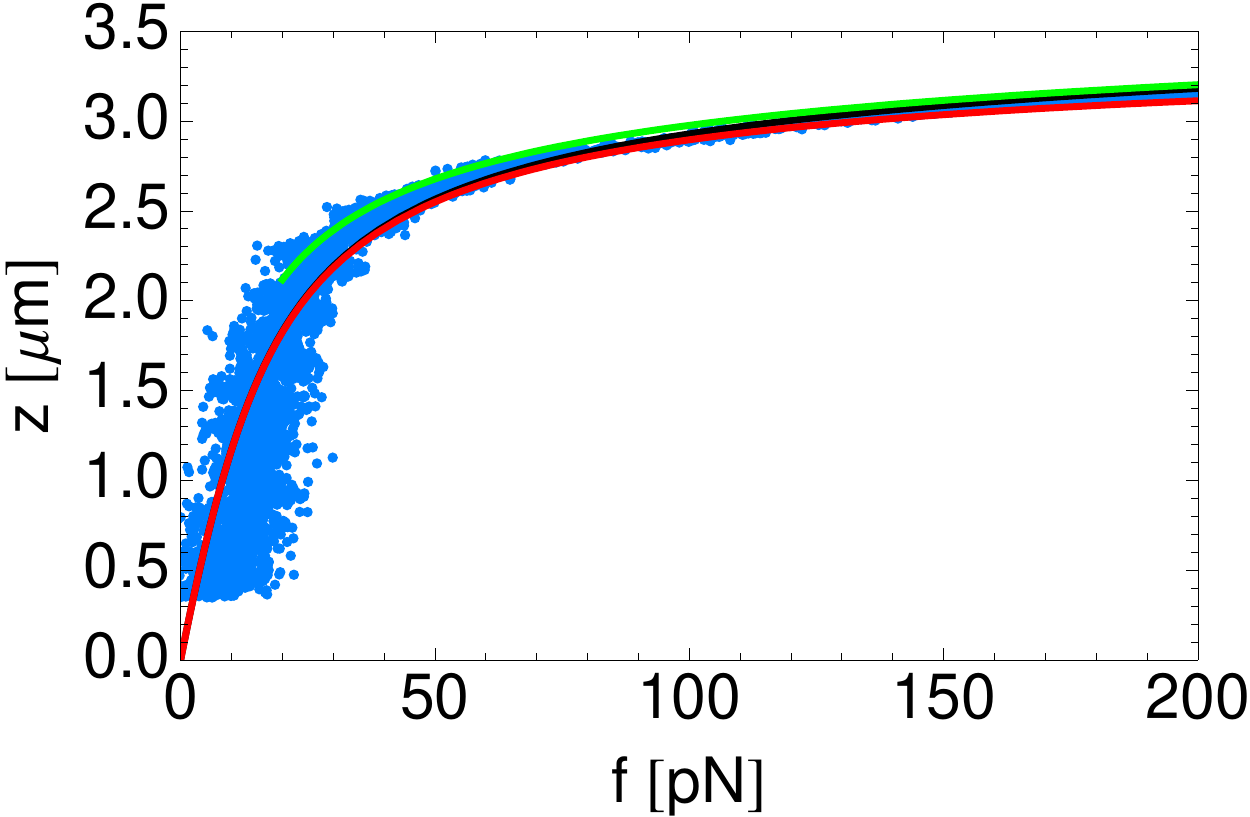}\\
\ \hfill {\bf (a)} \hfill\hfill {\bf (b)} \hfill \
\caption{{\bf (a)} Extension vs. force for a ssDNA. Data (symbols) are taken from H\"ugel \textit{et al.}~\cite{hugel}. The black solid curve corresponds to  a fit using the discrete worm like chain interpolation with the non-linear bond elasticity, \eq{interp} (the red one corresponds to discrete Marko Siggia interpolation \eq{interp1}). The parameters values are: $L=3.40 \;\mu$m, $\kappa=1.5$, $a=0.20$~nm. {\bf (b)} Zoom of (a) for small forces, together with the strong force limit \eq{ext_homo} (green).}
\label{figssDNA}
\end{figure}

The comparison with experimental data taken from~\cite{hugel} (kindly supplied by R.R. Netz) is shown in Fig.~\ref{figssDNA}(a) and (b). The fit using \eq{interp} is quantitatively good (the curve remains within the experimental error bars) and yields the following values for the fitting parameters: $L=3.40\;\mu$m, $\kappa=1.5$ and $a=0.20$~nm. This fit is very constrained since it is done on a very large range of forces, form 0 to 1200~pN. 

A first remark is that, although the persistence length is quite small $\ell_p=-a/\ln u(\kappa)\simeq0.24\;\mathrm{nm}\approx a$, the role of $\kappa$ is non-negligible in the low force regime and setting $\kappa=0$ (FJC model) leads to a poorer fit (data not shown).

As shown in Fig.~\ref{figssDNA}(b), the interpolation (black curve), \eq{interp} starts to deviate slightly from \eq{interp1} (red curve) for $f>150$~pN. Simply put, as already said by H\"ugel \textit{et al.}~\cite{hugel}, the non-linear stretching starts to play a significant role. Moreover, for $f> 400$~pN, we have $1/\sqrt{F^2+4\kappa F}\simeq F^{-1}<0.05$ which becomes smaller than $U_{\rm nl} (400 pN)=0.04$ in \eq{ext_stretch}. In other words, for $f> 400$~pN, the entropy becomes negligible, the influence of $\kappa$ and $a$ on the fit is small and the extension-curve is dominated by the non-linear stretching.

Although the $L$ and $\kappa$ values were expected, the effective monomer length $a=0.20$~nm is much smaller than the distance between two consecutive bases in DNA, $a_{ss}\approx0.7$~nm.  We define a corrective factor $b=a_{ss}/a=0.285$ which is the ratio between the expected ss base size $a_{ss}$ and the fitted monomer size value $a$. 
Note that Storm and Nelson~\cite{storm2} found similar values ($a=0.17$~nm) by using a Ritz variational approximation and fitting only on the $[0,200~\mathrm{pN}]$ range. Refs.~\cite{hugel,livadaru} focus on the non-linear elasticity which is significant for forces larger than 400~pN. By using a non-linear Freely Rotating Chain model at large forces they found a monomer size multiplied by 2 for polyvinylamin~\cite{livadaru,hugel} and a smaller monomer size by a factor 0.5 to 0.8 for peptides~\cite{hanke}. In our case, the \textit{actual} monomer length probed by a strong applied force is thus $1/b=3.5$ times smaller than the inter-base distance for ssDNA. In other words, the number of degrees of freedom $N$ for such strong forces increases by a factor 3.5.

This increase of $N$ at large forces raises an important question about a transition B-DNA to ssDNA in stretching force experiments. How to model this change using a mesoscopic model? This increase probably occurs abruptly during the transition and is likely to include chemical modifications unaccessible to classical mechanics.

\section{Analytical model for B to S-DNA transitions}
\label{sec_BtoSDNA}

While the previous section casts some doubt upon the adequacy of a mesoscopic model to model the B to ssDNA transition, we show in this section that it is possible to describe the B to S transition using such type of model. We use a Ising--Heisenberg coupled model, which has already been used by us for the theory of DNA denaturation~\cite{prl07,pre08}. Other works have used such types of models~\cite{storm1,storm2}. In this model, each base-pair $i$ is described by:  (1)~its normalized orientation $\mathbf{t}_i$ with $\Omega_i=(\theta_i,\varphi_i)$ the solid angle with respect to a fixed reference frame $(\hat x, \hat y, \hat z)$, (2)~its internal state $\sigma_i=\pm 1$ (corresponding to B or S base-pair internal state), and (3)~its length $a(\sigma_i)$. By generalizing \eq{HDWLC}, the effective Hamiltonian is 
\begin{equation}
\beta\mathcal{H} = \sum_{i=1}^{N-1} \kappa(\sigma_i,\sigma_{i+1})(1-\mathbf{t}_i\cdot\mathbf{t}_{i+1})- \beta a(\sigma_i) f \mathbf{t}_i\cdot\hat z+\mathcal{H}_I(\sigma_i,\sigma_{i+1})
\end{equation}

Compared to \eq{HDWLC}, the bending modulus $\kappa(\sigma_i,\sigma_{i+1})$ now depends on the internal state of base pairs $i$ and $i+1$ (we also note in the following $\mathbf{t}_i\cdot\mathbf{t}_{i+1}=\cos\gamma_{i,i+1}$) and the additional term 
\be
\mathcal{H}_I(\sigma_i,\sigma_{i+1}) = -J \sigma_i \sigma_{i+1} -\frac{\mu}2 (\sigma_i+\sigma_{i+1})
\ee 
is the internal Ising free energy associated with base pair $i$ and its interaction with base pair $i+1$ ($2\mu$ is the energy necessary to break one base-pair and $2J$ is the energy of a domain wall)~\cite{prl07,pre08}.

The transfer operator $\hat P$ is then defined by 
\bea
\langle \Omega_i,\sigma_i | \hat P| \Omega_{i+1},\sigma_{i+1} \rangle &=& a^2(\sigma)\exp\left[-\mathcal{H}_I(\sigma_i,\sigma_{i+1}) + \kappa(\sigma_i,\sigma_{i+1})(\cos\gamma_{i,i+1}-1) +\beta a(\sigma_i)f \cos \theta_i\right]\\
&=& a^2(\sigma)\exp\left[-\mathcal{H}_I(\sigma_i,\sigma_{i+1})+\frac{\kappa(\sigma_i,\sigma_{i+1})}2(\mathbf{t}_i-\mathbf{t}_{i+1})^2 +\beta a(\sigma_i)f t_{iz} \right]
\eea

\subsection{Exact diagonalization of $\hat P$}
\label{sec_BtoSDNA_exact}

The idea of dealing with the WLC model under forces in the spherical harmonics basis goes back to Marko and Siggia~\cite{Marko95}, even though their use in different but related fields of physics goes back to the 70s~\cite{Blume75}. In the present case, we diagonalize $\hat P$ by using the decomposition of a plane wave in spherical waves:
\bea
e^{\kappa\cos\gamma_{i,i+1}} =\sqrt\frac{\pi}{2\kappa} \sum_{\ell=0}^{\infty} I_{\ell+\frac12}(\kappa)\sum_{m=-\ell}^{\ell}Y_{\ell m}(\Omega_{i+1})Y^*_{\ell m}(\Omega_i)
\eea
which implies that $Y_{\ell m}$ is an eigenvector of $e^{\kappa\cos\gamma_{i,i+1}}$~\cite{Joyce67}:
\begin{equation}
\int \frac{{\rm d}\Omega_i}{4\pi} e^{\kappa\cos\gamma_{i,i+1}}Y_{\ell m}(\Omega_i)=
\sqrt\frac{\pi}{2\kappa} I_{\ell+\frac12}(\kappa)Y_{\ell m}(\Omega_{i+1}).
\label{vep:exp:cos}
\end{equation}
Note that the prefactor of $Y_{\ell m}$ in the rhs. term is the spherical Bessel function usually denoted by $i_\ell(\kappa)$ and that we also used the notation 
$e^{-\kappa} i_\ell(\kappa)=e^{-G_\ell(\kappa)}$ in Ref.~\cite{pre08}.

If $f=0$, $\hat P$ is block-diagonal in each $(\ell,m)$ subspace with matrix elements (we switch to lighter notations): 
\be
\lan \ell m, \sigma | \hat P | \ell' m', \sigma' \ran = a^2(\sigma)\exp\left[-G_\ell(\kappa(\sigma,\sigma'))-\mathcal{H}_I(\sigma,\sigma')\right]
\ee
They depend on $\ell$ but not on $m$. When diagonalizing each $2\times 2$ block, the eigenvalues are denoted by $\lambda_{\ell,\pm}$ and the eigenvectors by  $|\ell,\pm\rangle$. 
The partition function is $\mathcal{Z}=(4 \pi)^N \sum_{\pm,\ell}(2\ell+1) \lambda_{\ell,\pm}^N$ in case of periodic boundary conditions and by $\mathcal{Z}=(4 \pi)^N \sum_{\ell,\pm} 
\langle V | 0,\pm \rangle^2 \lambda_{\ell,\pm}^N$ in case of free ends (where $|V\ran$ is the adequate free-end vector)~\cite{pre08}. Of course, boundary conditions are irrelevant at large $N$~\cite{Marko10}, and we have checked that finite size effects are negligible for the DNAs studied in the experiments considered in the following.

If $f \neq 0$, $\hat P$ is block-diagonal, but blocks are now infinite because different values of $\ell$ are coupled. A block thus corresponds to a given value of $m$. \eq{vep:exp:cos} has to be adapted to this case~\cite{Marko04,Marko05,Rahi08,Marko10} (Note that contrary to~\cite{Rahi08}, we do not need to explicitly treat  the two single strands in the B to S transition). In Dirac notations, we have:
\be
\langle \ell m| e^{\kappa (\cos \gamma-1)+F\cos \theta}  | \ell' m' \rangle =  4\pi (-1)^m \delta_{m,m'} \sqrt{(2\ell+1)(2\ell'+1)}  e^{-\kappa}i_{\ell'}(\kappa)\sum_{\ell_1=0}^\infty (2\ell_1+1)
\w3j{\ell_1}{\ell}{\ell'}000\w3j{\ell_1}{\ell}{\ell'}0m{-m} i_{\ell_1}(F)
\label{wigner}
\ee
where we have used Wigner 3j-symbols (and $m\leq \ell,\ell'$). Such use of Wigner 3j-symbols already appeared in Blume et al.~\cite{Blume75}. In this expression, $\kappa$ means $\kappa(\sigma,\sigma')$ and $a$ means $a(\sigma)$. In practice, to diagonalize each infinite block, a cutoff $\lmax\approx 8$ on $\ell$ must be chosen (large values of $\ell$ do not give significantly different results). Once $\hat P$ is diagonalized, $\mathcal{Z}$ can be computed from which the extension $z$ is derived.

\subsection{Strong force approximation}
\label{sec_BtoSDNA_sfa}

Similarly to Section~\ref{sec_ssDNA}, we use now the description of $\hat P$ in tangent vectors $\bt_i$. An extension of the spinor eigenvector equation is now, in a symmetric form:
\be
\int \md \bt' 
\left({\begin{array}{*{20}c}
   {a_+^2e^{\mu+J} \;\bK_{++}(\bt,\bt') } & {a_+a_-e^{-J} \;\bK_{+-}(\bt,\bt')}  \\
   {a_-a_+e^{-J} \;\bK_{-+}(\bt,\bt') } & {a_-^2e^{-\mu+J} \;\bK_{--}(\bt,\bt')}  \\
\end{array}} \right)
\left({\begin{array}{*{20}c}
   \Phi_{+}(\bt')  \\
   \Phi_{-}(\bt')  \\
\end{array}} \right) 
=\Lambda 
\left({\begin{array}{*{20}c}
   \Phi_{+}(\bt)  \\
   \Phi_{-}(\bt)  \\
\end{array}} \right) 
 \label{TM}
\ee
where $\Lambda$ is the eigenvalue and  $\Phi_\sigma(\bt)=\phi_{x\sigma}(t)\phi_{y\sigma}(t)$ are the unknown eigenfunctions. The transfer operator $\bK_{\sigma\sigma'}(\bt,\bt')=K_{\sigma\sigma'}(t_x,t_x')K_{\sigma\sigma'}(t_y,t_y')$ with $\sigma,\sigma'=\pm$ is the generalization of \eq{TOH} where 
\be
K_{\sigma\sigma'}(t,t')= \exp\left[-\frac12\kappa(\sigma,\sigma')(t-t')^2-\frac14F(\sigma)t^2-\frac14F(\sigma')t'^2\right]
\label{TM2}
\ee
Searching for an exact diagonalization of the transfer matrix is difficult because a Gaussian wave such as in Section~\ref{sec_ssDNA} is no more an exact eigenfunction. This is due to the fact that cross-terms are not symmetric in $t$ and $t'$: $K_{1,-1}(t,t')\neq K_{1,-1}(t',t)$ [but $K_{1,-1}(t,t')= K_{-1,1}(t',t)$]. For the same reason, the Ritz variational scheme~\cite{storm1,storm2}  does not work for this model with a Gaussian variational eigenfunction.

Nevertheless,  eigenfunctions are Gaussian in 3 limits: 1) the homogeneous chain (this has been proved in Section~\ref{sec_ssDNA}), 2) the zero force limit $f=0$, and 3) in the freely jointed chain limit, $\kappa_i=0$. Indeed, by inserting in \eq{TM} Gaussian wave functions $\phi_{\sigma}=C_\sigma \exp(-\alpha_{\sigma} t^2/2)$ and using \eqs{kernel1}{result1}, we find
\bea
C_+ \sqrt{\frac{2\pi}{\kappa_{++} + \frac{F_+}2+\alpha_+}}  e^{-\alpha_{++} t^2/2}+C_- \sqrt{\frac{2\pi}{\kappa_{+-} +\frac{F_-}2+\alpha_-}} e^{-\alpha_{+-} t^2/2} &=& \lambda e^{-\alpha_+ t^2/2}\label{Cp} \\
C_+ \sqrt{\frac{2\pi}{\kappa_{-+}+\frac{F_+}2+\alpha_+}} e^{-\alpha_{-+} t^2/2}+C_- \sqrt{\frac{2\pi}{\kappa_{--}+\frac{F_-}2+\alpha_-}} e^{-\alpha_{--} t^2/2} &=&\lambda e^{-\alpha_- t^2/2} \label{Cm} 
\eea
where
\be
\alpha_{\sigma\sigma'}=\kappa_{\sigma\sigma'}+\frac{F_\sigma}2-\frac{\kappa_{\sigma\sigma'}^2}{\kappa_{\sigma\sigma'}+\frac{F_{\sigma'}}2+\alpha_{\sigma'}}
\ee
By assuming $\alpha_{\pm\pm}=\alpha_{\pm}=\sqrt{(F_\pm/2)^2+\kappa_{\pm\pm}F_\pm}$, and dividing \eqs{Cp}{Cm} respectively by $\exp(-\alpha_{\pm} t^2/2)$, one finds
\bea
C_+ \sqrt{\frac{2\pi}{\kappa_{++} + \frac{F_+}2+\alpha_+}} +C_- \sqrt{\frac{2\pi}{\kappa_{+-} +\frac{F_-}2+\alpha_-}} e^{(\alpha_{+}-\alpha_{+-}) t^2/2} &=& \lambda  \label{G1}\\
C_+ \sqrt{\frac{2\pi}{\kappa_{-+}+\frac{F_+}2+\alpha_+}} e^{(\alpha_--\alpha_{-+}) t^2/2} +C_- \sqrt{\frac{2\pi}{\kappa_{--}+\frac{F_-}2+\alpha_-}} &=&\lambda \label{G2}
\eea
It is then straightforward to check that in case (2) ($f=0$), we have $\alpha_{\pm\mp}=0=\alpha_\pm$ and  in case (3) ($\kappa_{\sigma\sigma'}=0$) $\alpha_{\pm\mp}=F_\sigma/2=\alpha_\pm$ and the Gaussians of the cross-terms are equal to 1.

To proceed further, we make the approximation that the Gaussians in \eqs{G1}{G2} are negligible and thus assume that they are equal to 1 for all the parameter values. This allows us to write an effective Ising model since  \eqs{G1}{G2} do not depend on $t$ anymore. Hence, coming back to \eq{TM}, the transfer matrix reduces to a simpler $2\times2$ Ising transfer matrix:
\be
\hat P_{\rm eff} = e^{\Gamma _0} \left(
{\begin{array}{*{20}c}
   {e^{\mu_0  + J_0 } } & {e^{ - J_0 -\delta} }  \\
   {e^{ - J_0 +\delta} } & {e^{ - \mu_0  + J_0 } }  \\
\end{array}} \right)\\
\ee
with force- and temperature-dependent Ising parameters (we switch to the subscripts $B$ for B-DNA instead of $+$, and $S$ instead of $-$ for S-DNA):
\bea
\mu_0 &=& \mu-\ln\gamma+F\frac{1-\gamma}2+\frac12\ln\left(\frac{\kappa_S+F\gamma/2+\alpha_S}{\kappa_B+F/2+\alpha_B}\right) \label{L0}\\
J_0 &=& J+\frac14\ln\left[\frac{(\kappa_{SB}+F/2+\alpha_B)(\kappa_{SB}+F\gamma/2+\alpha_S)}{(\kappa_{B}+F/2+\alpha_B)(\kappa_{S}+F\gamma/2+\alpha_S)}\right] \label{J0}\\
\Gamma_0 &=& \ln(2\pi\gamma) + F\frac{1+\gamma}2-\frac14\ln\left[\left(\kappa_{SB}+\frac F2+\alpha_B\right)\left(\kappa_{SB}+\frac{F\gamma}2+\alpha_S\right)\left(\kappa_{B}+\frac F2+\alpha_B\right)\left(\kappa_{S}+\frac{F\gamma}2+\alpha_S\right)\right] \label{Gamma0}\\
\delta &=& \frac12 \ln\left(\frac{\kappa_{SB}+F/2+\alpha_S}{\kappa_{SB}+\gamma F/2+\alpha_B}\right) \label{delta}
\eea
where
\be
\gamma \equiv \frac{a_S}{a_B} \qquad \alpha_B=\sqrt{\kappa_BF +\left(\frac{F}2\right)^2}\qquad\alpha_S=\sqrt{\kappa_S\gamma F +\left(\frac{\gamma F}2\right)^2}
\ee

In the limit $N\to\infty$, the partition function is then given by $\mathcal{Z}=\Lambda^N$ where the largest eigenvalue of the Ising matrix is~\footnote{The parameter $\delta$ does not enter the eigenvalues but slightly changes the eigenvectors compared to the true Ising problem, which is negligible in the $N\to\infty$ limit.} 
\be
\Lambda = e^{\Gamma_0+J_0}\left(\cosh \mu_0+\sqrt{\sinh^2\mu_0+e^{-4J_0}}\right)
\ee
The ``magnetization'' and the two-point correlation of this effective Ising model are (see Ref.~\cite{pre08}):
\bea
\lan\sigma_i\ran &=& \frac{\sinh \mu_0}{\sqrt{\sinh^2\mu_0+e^{-4J_0}}} \label{magnet}\\
\lan\sigma_i\sigma_{i+1}\ran &=& \lan\sigma_i\ran^2+\left(1-\lan\sigma_i\ran^2\right)\frac{\cosh \mu_0-\sqrt{\sinh^2\mu_0+e^{-4J_0}}}{\cosh \mu_0+\sqrt{\sinh^2\mu_0+e^{-4J_0}}}\label{corr}
\eea
where \eq{magnet} yields the fraction of base-pairs in the B (resp. S) state 
\be
\varphi_{B,S}(F)=\frac{1\pm \lan\sigma_i\ran}2
\label{phiS}
\ee 
as a function of the force. The extension computed according to \eq{def_ext} is thus
\be
\frac{z}{a_B N} = \left(1-\frac1{2\alpha_B}\right)\varphi_B+\gamma\left(1-\frac1{2\alpha_S}\right)\varphi_S +\frac{\lan\sigma_i\sigma_{i+1}\ran-1}4\left(\frac1{2\alpha_B}\frac{\kappa_B-\kappa_{BS}}{\kappa_{BS}+F/2+\alpha_B}+\frac\gamma{2\alpha_S}\frac{\kappa_S-\kappa_{BS}}{\kappa_{BS}+\gamma F/2+\alpha_S}\right)
\label{ext}
\ee
which shows that the last term is only relevant  close to the transition where $\lan\sigma_i\sigma_{i+1}\ran\neq1$.

\eq{ext} is the second important result of the paper. As it is constructed, this formula is an interpolation between several limits. 
First, the result for an homogeneous chain in the strong force approximation, \eq{ext_homo}, is recovered by setting $\kappa_B=\kappa_S=\kappa_{BS}$ and $\gamma=1$ in \eq{ext}. 
Second, in the zero force limit $F=0$, \eqs{L0}{delta} simplify to
\be
\mu_0 = \mu+\frac12\ln\left(\frac{\kappa_S}{\gamma^2\kappa_B}\right) \qquad
J_0 = J+\frac14\ln\left(\frac{\kappa_{SB}^2}{\kappa_{B}\kappa_{S}}\right) \qquad
\Gamma_0 =\frac14\ln(\kappa_{SB}^2\kappa_{B}\kappa_{S})-\ln(2\pi\gamma) \qquad \delta=0
\ee
which are the renormalized Ising parameters already found in \cite{jpcm09}.
The FJC model corresponds to $\kappa_B=\kappa_S=\kappa_{BS}=0$, and \eq{ext} reduces to
\be
\frac{z}{a_B N} = (\varphi_B+\gamma\varphi_S)-\frac1F
\ee
which is \eq{fjc} slightly modified to take into account to the two accessible values for the base pair length $a_i$.

Finally, far from the transition, defined as $\lan\sigma_i\ran =0$ or equivalently $\mu_0(f_c)=0$ for infinitely long DNAs, that is for forces such that $\lan\sigma_i\sigma_{i+1}\ran =1$, we find B- (or S-) stretching behaviour, for $\mu_0>0$ (respectively $\mu_0<0$):
\bea
\frac{z}{a_B N} \simeq 1-\frac1{\sqrt{4\kappa_B F + F^2}} \qquad f &\ll& f_c \label{lowF}\\
\frac{z}{a_B N} \simeq \gamma-\frac\gamma{\sqrt{4\kappa_S \gamma F + \gamma^2F^2}} \qquad f &\gg& f_c\label{highF}
\eea
This last result is identical to \eq{ext_homo} provided that the extension and the force are renormalized by the S-monomer length $a_S=\gamma a_B$. \eqs{lowF}{highF} are a generalization of the result of Cizeau and Viovy~\cite{cizeau} where the continuous Marko-Siggia interpolation, valid for range $1/\kappa\ll F\ll 4\kappa$, was used.
\\

In experiments, large forces on the order of several hundreds of picoNewtons are applied to B-DNA such that helix stretching occurs. This stretching is related to the torsional elasticity of the double helix, such as for a helical spring. It is incorporated linearly, following \eq{ext_stretch}, by replacing $f$ by $f(1+f/E_B)$~\cite{Marko95,storm1,storm2}  in the prefactors independent of $t$ and $t'$ in the matrix elements of \eq{TM}, where $E_B$ (in pN) and the related adimensional $\tilde E_B=\beta a_B E_B$ is the stretching modulus in the B state, taken as a fitting parameter. \eq{ext} becomes~\footnote{Contrary to~\cite{storm1,storm2}, within our discrete chain model, we do not need to consider any stretching modulus in the S form to fit accurately the data. Assuming that the S-form is unstacked and unwound, the stretching modulus is expected to be close to the ssDNA one, on the order of $10^4$~pN [see \eq{UNL}], and is negligible in this range of forces.}
\bea
\frac{z}{a_B N} = \left(1+\frac{F}{\tilde E_B}-\frac1{2\alpha_B}\right)\varphi_B+\gamma\left(1-\frac1{2\alpha_S}\right)\varphi_S +\frac{\lan\sigma_i\sigma_{i+1}\ran-1}4\left(\frac1{2\alpha_B}\frac{\kappa_B-\kappa_{BS}}{\kappa_{BS}+F/2+\alpha_B}+\frac\gamma{2\alpha_S}\frac{\kappa_S-\kappa_{BS}}{\kappa_{BS}+\gamma F/2+\alpha_S}\right)
\label{extbond}
\eea
\eqs{ext}{extbond} allow us to fit with a high accuracy the various experimental DNA stretching curves taken from the literature and also compare perfectly well with the semi-analytical exact formula, \eq{wigner}, as illustrated in the next Section.

\section{Comparison with experimental force-extension curves}
\label{sec_BtoSDNA_comp}

We now compare our theoretical approach to various experimental data. Some of them, coming from optical tweezers experiments, are extracted from~\cite{storm2}. Rief \textit{et al.}~\cite{rief} also conducted several experiments using AFM on $\lambda$-phage DNA, poly(dG-dC) and poly(dA-dT) DNAs, in order to explore the role of base-sequence on DNA stretching. 

\subsection{Overstretching transition for poly(dG-dC) and $\lambda$-DNA around 60--80~pN}
\begin{figure}[t]
\includegraphics[width=0.48\linewidth]{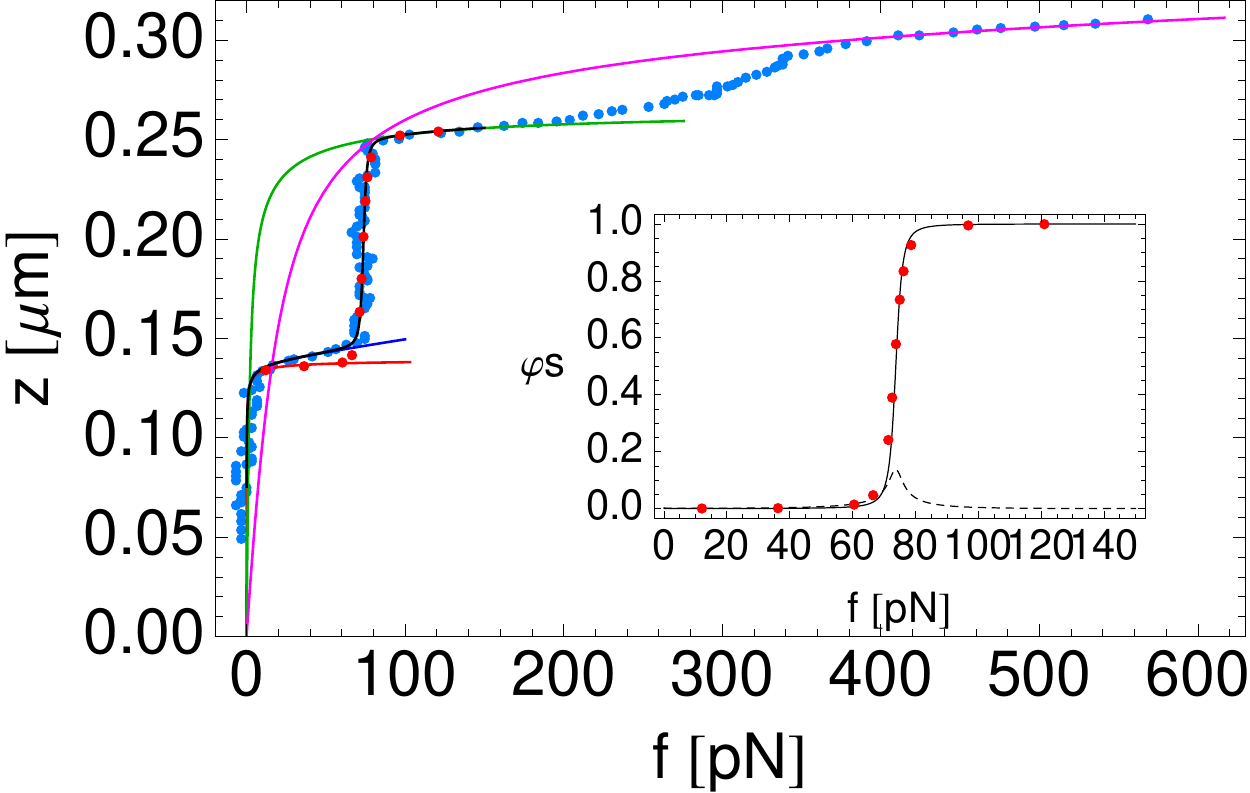}\hfill\includegraphics[width=0.48\linewidth]{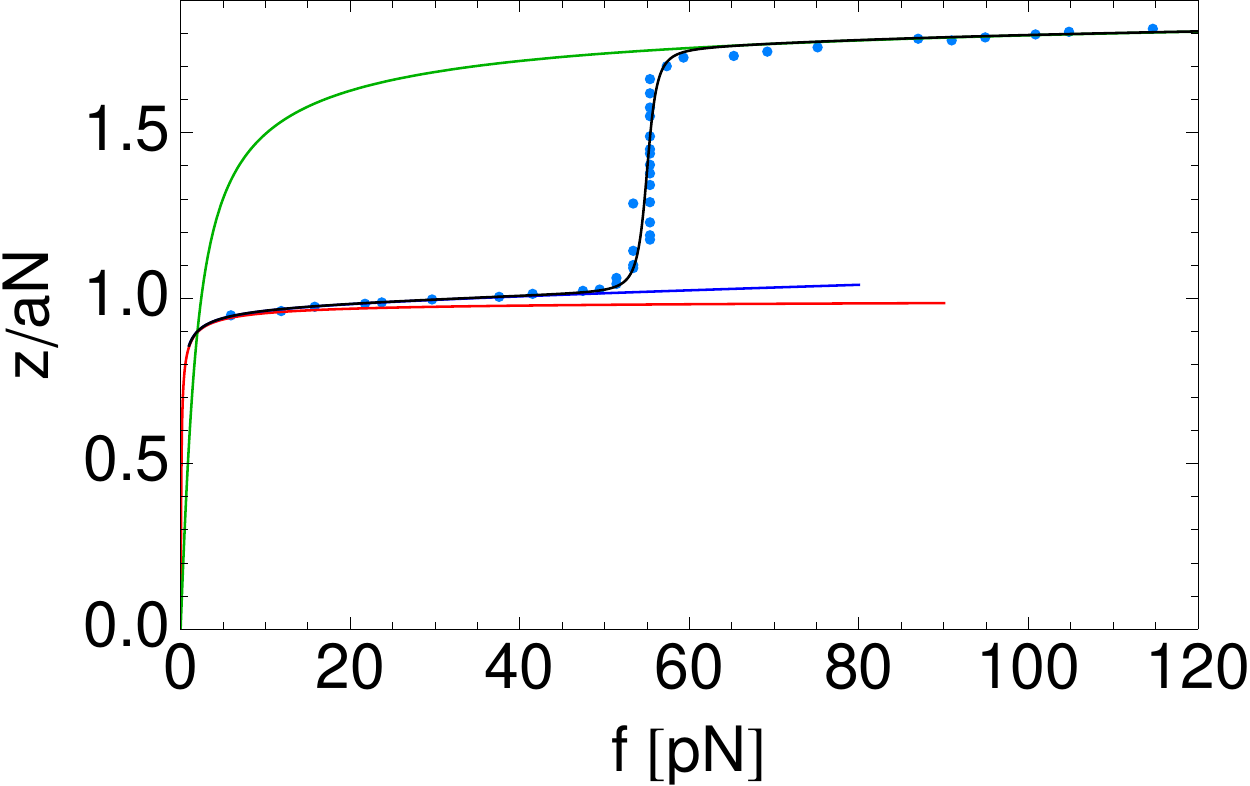}\\
\ \hfill {\bf (a)} \hfill\hfill {\bf (b)} \hfill \
\caption{{\bf (a)} Extension vs. force for a poly(dG-dC). Data (blue symbols) are taken from Rief \textit{et al.}~\cite{rief}. Solid curves correspond to the discrete Discrete Worm like chain interpolation, \eq{interp1}, for B-DNA (red), S-DNA (green) and with non-linear extensibility, \eq{interp}, for ssDNA (pink). The black curve corresponds to \eq{extbond}, where linear stretching is included as shown by the blue curve for pure B-DNA. The red symbols correspond to the semi-analytical calculation \eq{wigner} with $\ell_{\max}=8$. Parameters values are: $L_B=0.14 \;\mu$m, $\kappa_B=147$, $\gamma_1=1.89$, $\kappa_{S}=\kappa_{BS}=3.8$, $\gamma_2=1.145$,  $b=0.285$, $\kappa_{ss}=5.54/(2\gamma_1\gamma_2)=1.28$, $\mu=4.5$, $J=1.7$, $E_B=1200$~pN. {\bf Inset}: Fraction of base-pairs in the S state vs. force, \eq{phiS} and Ising correlation function $1-\lan\sigma_i \sigma_{i+1}\ran$ defined in \eq{corr} (dashed curve). {\bf (b)}  Same as (a) for $\lambda$-phage DNA. Data (blue symbols) are data taken from~\cite{rief}. Parameters values are: $\kappa_B=147$, $\gamma=1.88$, $\kappa_{S}=\kappa_{BS}=4$, $\mu=3.85$, $J=2.05$, $E_B=1400$~pN.}
\label{figrief}
\end{figure}

\begin{figure}[t]
\includegraphics[width=0.48\linewidth]{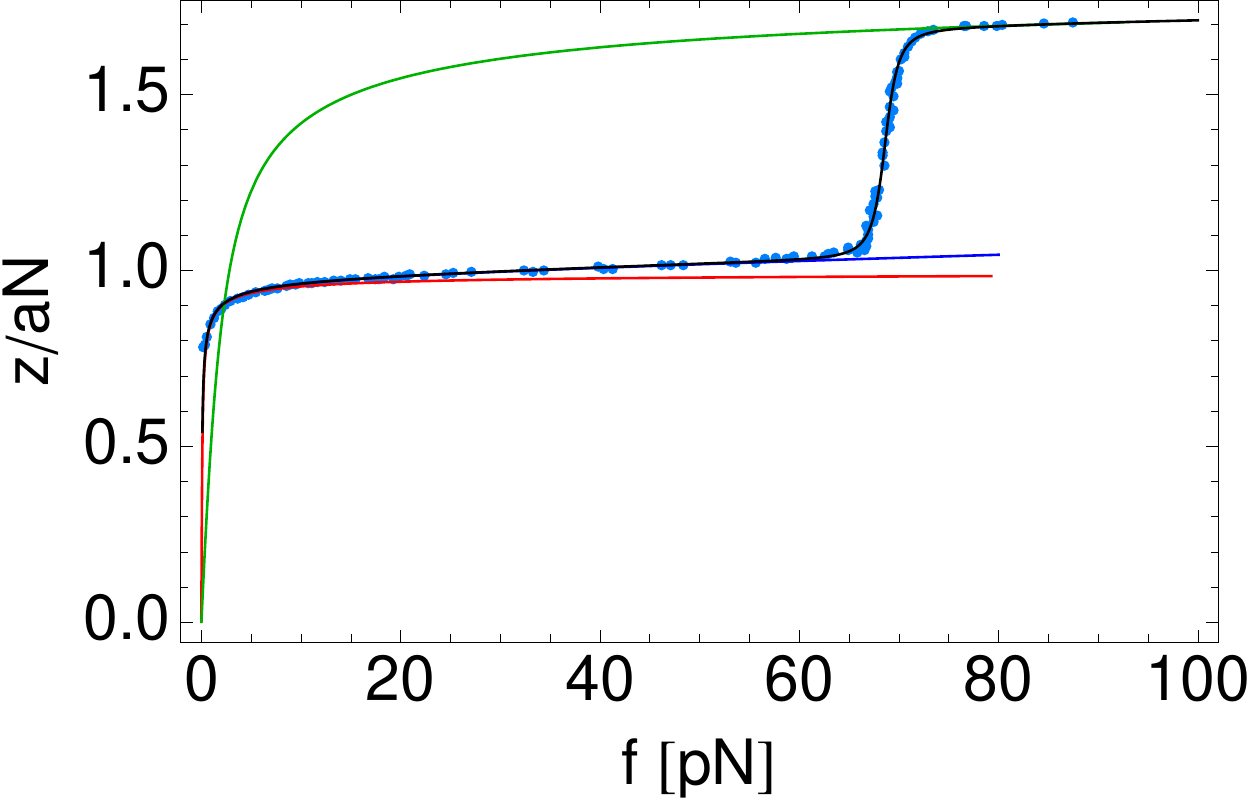}\hfill\includegraphics[width=0.47\linewidth]{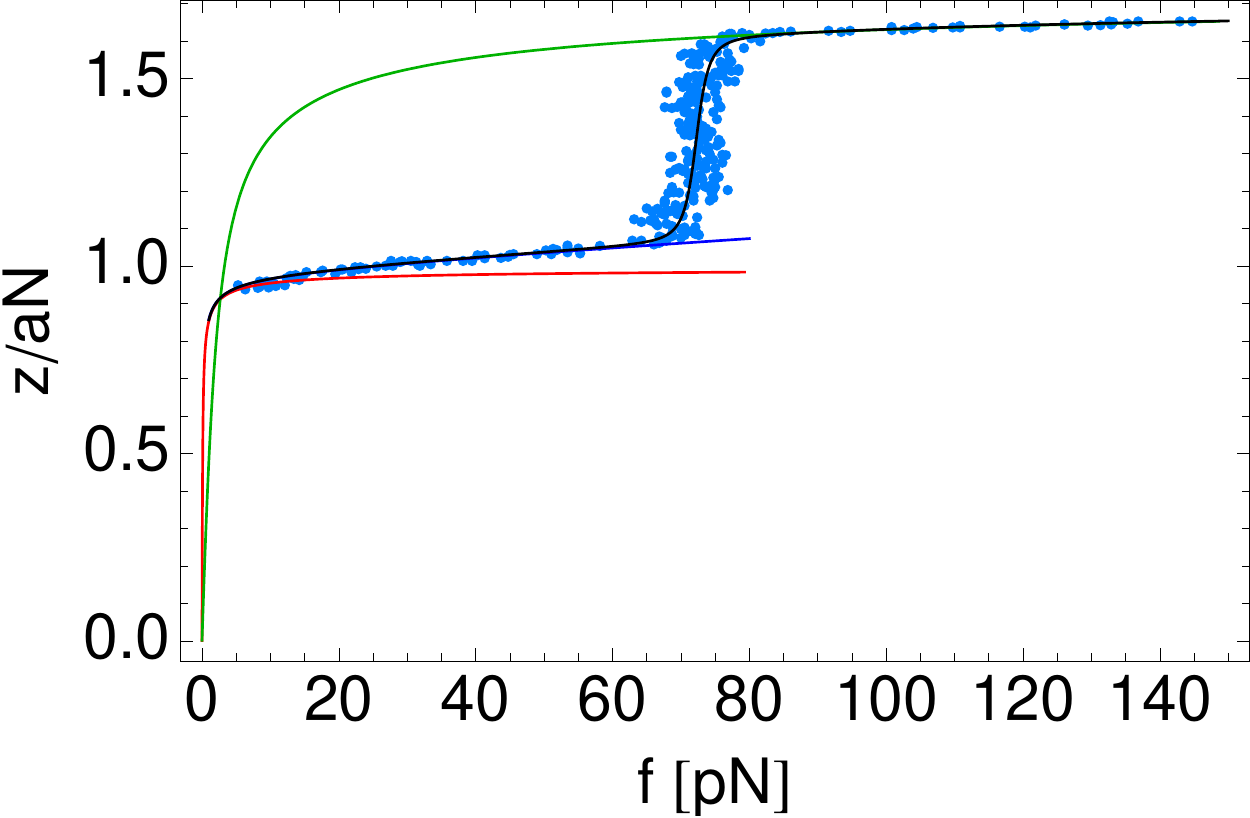}\\
\ \hfill {\bf (a)} \hfill\hfill {\bf (b)} \hfill \
\caption{Extension vs. force for a $\lambda$-phage DNA. Data (symbols) are due to Cui and Bustamante for (a) and Cluzel \textit{et al.}~\cite{Cluzel96} for (b), and both taken from Storm and Nelson~\cite{storm2}. The solid curves correspond to the discrete Discrete Worm like chain interpolation, \eq{interp}, for B-DNA without stretching (red), with stretching (blue) and ssDNA (green). The black curve correspond to \eq{extbond}, where linear stretching is included. Parameters values: $\kappa_B=147$, $\kappa_{S}=\kappa_{BS}=4$, $1/E_{S}=0$, and {\bf (a)} $\gamma=1.795$,  $\mu=4.015$, $J=2$, $E_B=1300$~pN; {\bf (b)} $\gamma=1.715$, $\mu=3.85$, $J=1.9$, $E_B=880$~pN.  }
\label{lambdaSN}
\end{figure}

First, we focus on the B to S transition which occurs for $f=f_c\simeq 60-80$~pN with small differences related to the DNA sequence. We compared our analytical result \eq{extbond} to experiments made on polyGC taken from~\cite{rief} [see Fig.~\ref{figrief}(a)] and $\lambda$-phage DNA [figs.~\ref{figrief}(b) and Fig.~\ref{lambdaSN}], and  \eq{extbond} leads to very good fits of experimental data. The fitting procedure is detailed in the appendix.

To begin with, we have checked in Fig.~\ref{figrief}(a) that the semi-analytical calculation using the result of the Section IIIA (red symbols) and the strong force approximation of Section IIIB (black solid curve) are superimposed. Note that the results of Section IIIA are done without linear elasticity for B-DNA. This is the reason why there is a slight difference before the transition since we have plotted only \eq{extbond} and not \eq{ext} for sake of clarity. It proves the validity of the strong force approximation  used to derive \eq{extbond} for the B to S transition. 
This is due to the fact that the transition occurs in the force range (60--80~pN) where $f\gg\bar f_B=\kB T/(a_{B}\kappa_B)\simeq 0.1$~pN and $f\gg\bar f_S=\kB T/(a_{S}\kappa_S)\simeq 2$~pN.

This is also confirmed by the plot of $\varphi_S(f)$ in the inset of Fig.~\ref{figrief}(a), where both results are superimposed. Moreover, the superimposition of \eq{extbond}, valid for $N\to\infty$, and the results of Section IIIA computed for $N\simeq 700$ provide the undisputed evidence that the very small correction due to finite $N$ lies within error bars. Note that the transition is abrupt as shown by the plot of $1-\lan\sigma_i\sigma_{i+1}\ran$ in Fig.~\ref{figrief}(a), and \eq{extbond} gives the two correct limits far from the transition \eqs{lowF}{highF}.
In the fitting procedure, the parameter $\kappa_{BS}$ plays a similar role as the parameter $J$ (see \eq{J0}). This is the reason why we chose $\kappa_{BS}=\kappa_{S}$.

Furthermore, we notice that the fits of Figs.~\ref{figrief} and~\ref{lambdaSN} yield similar values for $\gamma$, between 1.72 and 1.89, which are also comparable to those obtained by Storm and Nelson~\cite{storm1,storm2} for $\lambda$-DNA (1.7--1.8). While the bending modulus of B-DNA is taken to  be $147\,\kB T$, the S-DNA one is much smaller, between 3.8 and $4\, \kB T$. 
Finally, similarly to~\cite{storm1,storm2} we find $E_B\approx 1000$~pN. Contrary to Refs.~\cite{storm1,storm2}, we do not need to introduce an linear modulus for the S-form, which can be attributed to the fact that a continuous WLC model was used in~\cite{storm1,storm2}, instead of a discrete one.

By fitting the transition using \eq{extbond}, one finds $3.85<\mu<4.5$ in $\kB T$ units, which are reasonable values compared to that extracted from the Poland-Scheraga model~\cite{polscher} and fits of denaturation curves~\cite{wartben}. It is roughly twice the value found for for poly(dA)-poly(dT) in~\cite{pre08,jpcm09}, and is consistent with the fact that GC base-pairing energy is larger than AT one. The cooperativity parameter $J$ is $1.7<J<2.05$ which is also reasonable and a little smaller than the value of 3.6 found for poly(dA)-poly(dT) in~\cite{pre08,jpcm09}. The small variation of the fitting parameter values from sample to sample are probably due to the slight differences in DNA sequences and salt conditions. In Fig.~\ref{polyGC2dtrans}(a) is plotted, in the $(T, f)$ plane, the coexistence line defined by setting $\mu_0 (T_m,F_c)=0$ in \eq{L0} with the parameters values of Fig.~\ref{figrief}(a). It shows the same behaviour as in Refs.~\cite{Rahi08,Netz10}, with a re-entrance for (unreachable) high temperatures, and decreases linearly in the accessible temperature window~\cite{zhang}.

To conclude this Section, the fact that \eq{extbond} allows us to fit the transition observed experimentally for poly(dG-dC) and $\lambda$-DNA indicates that the second state is indeed a S state and not a ss state. Indeed, a good fit of a transition to a ssDNA state around 60--80~pN would impose a much smaller value of the monomer size as discussed in detail in Section~\ref{sec_ssDNA} and below.

\subsection{Second transition for poly(dG-dC) around 350~pN}

\begin{figure}[t]
\includegraphics[width=8.2cm]{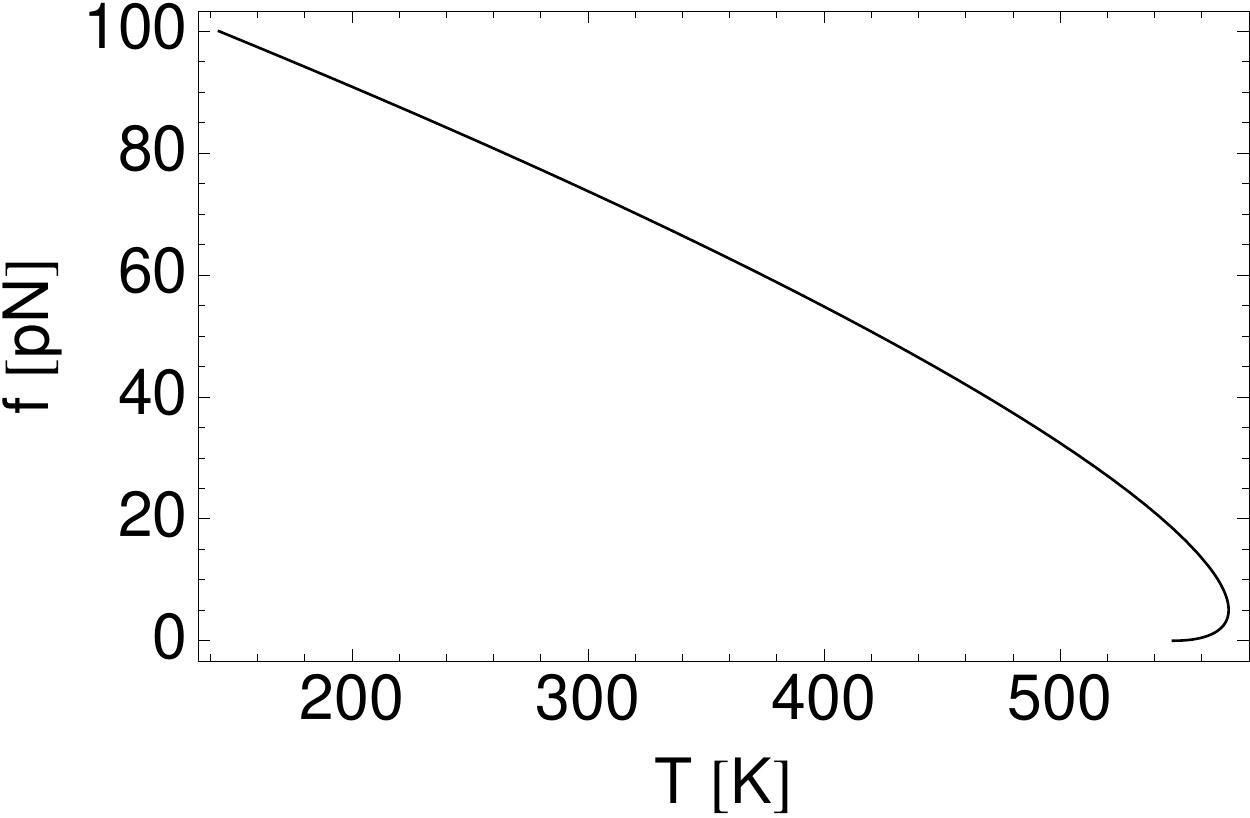}\hfill
\includegraphics[width=8.5cm]{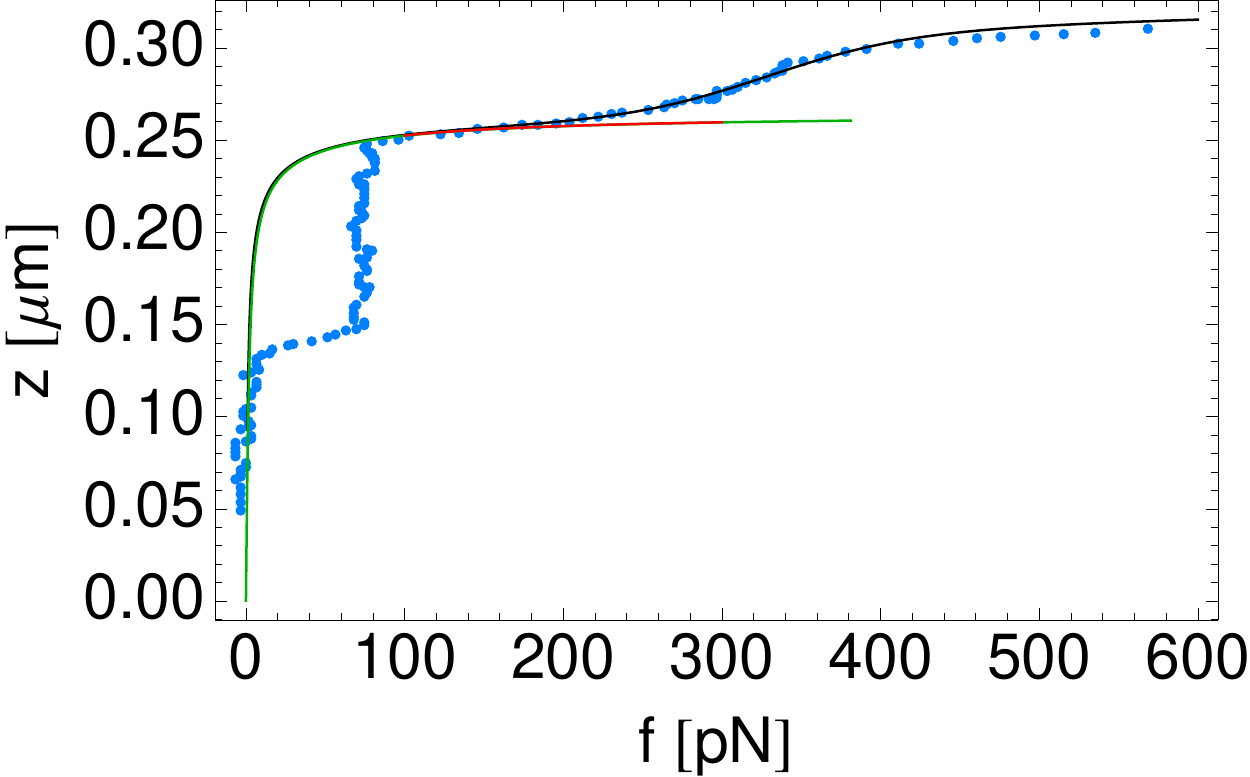}\\
\ \hfill {\bf (a)} \hfill\hfill {\bf (b)} \hfill \
\caption{{\bf (a)} Phase diagram in the temperature--force space corresponding to the first transition of Fig.~\ref{figrief}(a).  {\bf (b)} Fit (black curve) of the second transition for poly(dG-dC) using \eq{extbond} where the linear stretching term is replaced by the non-linear stretching one for ssDNA given by \eq{UNL}. The green solid curve is the same as in Fig.~\ref{figrief}(a) and the red one is the large force limit \eq{ext_homo}. The parameters are the same as in  Fig.~\ref{figrief}(a), $Na_S=Na_B\gamma_1=0.265 \;\mu$m, $\kappa_S=3.8$, $\gamma_2=1.145$, $\kappa_{ss}=\kappa_{Sss}=1.28$; and the fitting parameters are $\mu=3.8$ and $J=0$. Note that, compared to  Fig.~\ref{figrief}(a), the fit is poorer for ssDNA since no parameter $b$ is introduced.}
\label{polyGC2dtrans}
\end{figure}

Rief \textit{et al.}~\cite{rief} observed a second transition when they stretched a poly(dG-dC) at larger forces, around $\simeq350$~pN, as shown in Fig.~\ref{polyGC2dtrans}(b). They argued that this transition corresponds to a S to ssDNA transition, where the final state corresponds to one single ssDNA strand  which remains tethered, the second strand being unpeeled~\cite{rief}.  Indeed, this second transition, which is very smooth between, roughly 200 and 400~pN,  shows an hysteresis which varies with the applied pulling speed of the AFM tip. They also observed this second transition for $\lambda$-DNA at a smaller force, $f_c\simeq150$~pN.

Following these arguments, we try to model this second transition. First we use the result of Section~\ref{sec_ssDNA}, where the value of the actual bond size is $a=0.20$~nm. We then fit the experimental data at strong forces in Fig.~\ref{figrief}(a). One finds consistently $L_{ss}=\gamma_1\gamma_2 L_{B}= 2.16 L_{B}$ and $\kappa_{ss}=1.23\simeq 5.54 /(2\gamma_1\gamma_2)$, where $\gamma_1=a_S/a_B$ and $\gamma_2=a_{ss}/a_S$ and the value 5.54 is taken from~\cite{prl07,pre08}. These two values corresponds to the distance between two adjacent base-pairs along the helix ($\approx0.7$~nm) and to the accepted bending modulus value of a single ssDNA strand (persistence length $\ell_p\approx1$~nm~\cite{Smith96}). Note that the persistence length of ssDNA can vary a lot with the salt concentration~\cite{mano}.

Second, to fit the transition, we use \eq{extbond} where the B state becomes the S one and the S state is the ss one. However, as explained in Section~\ref{sec_ssDNA}, the change of degrees of freedom from B to ssDNA should prevent the success of the fit. But, since the ssDNA form is obtained for $f>400$~pN, following Section~\ref{sec_ssDNA}, the entropy is not dominant for this force range, the stretching being essentially due to bond deformations modeled by the non-linear stretching \eq{UNL}. Hence, in the absence of any model with different degrees of freedom in S and ss states, it is reasonable, as a first attempt, to keep the same number of degrees of freedom for this case [see in Fig.~\ref{polyGC2dtrans}(b)]: $a=a_{ss}=\gamma_1\gamma_2a_B=2.16a_B$ (or $b=1$).

Within this hypothesis, we are able to fit approximatively the experimental curve by replacing  the linear stretching term, $F/\tilde E_B$, in \eq{extbond} by the non-linear stretching one for ssDNA given by \eq{UNL}. The values of the fitting Ising parameters for this transition are $\mu=3.8$ and $J=0$. It indicates that this transition is not cooperative at all. This is consistent with the commonly accepted picture of a destacked S-DNA: during the S to ssDNA transition, only the breaking of the hydrogen bonds between base-pairs occurs, the aromatic rings being already destacked in the S state.

\subsection{Nature of the transition for poly(dA-dT)}

\begin{figure}[t]
\includegraphics[width=8.5cm]{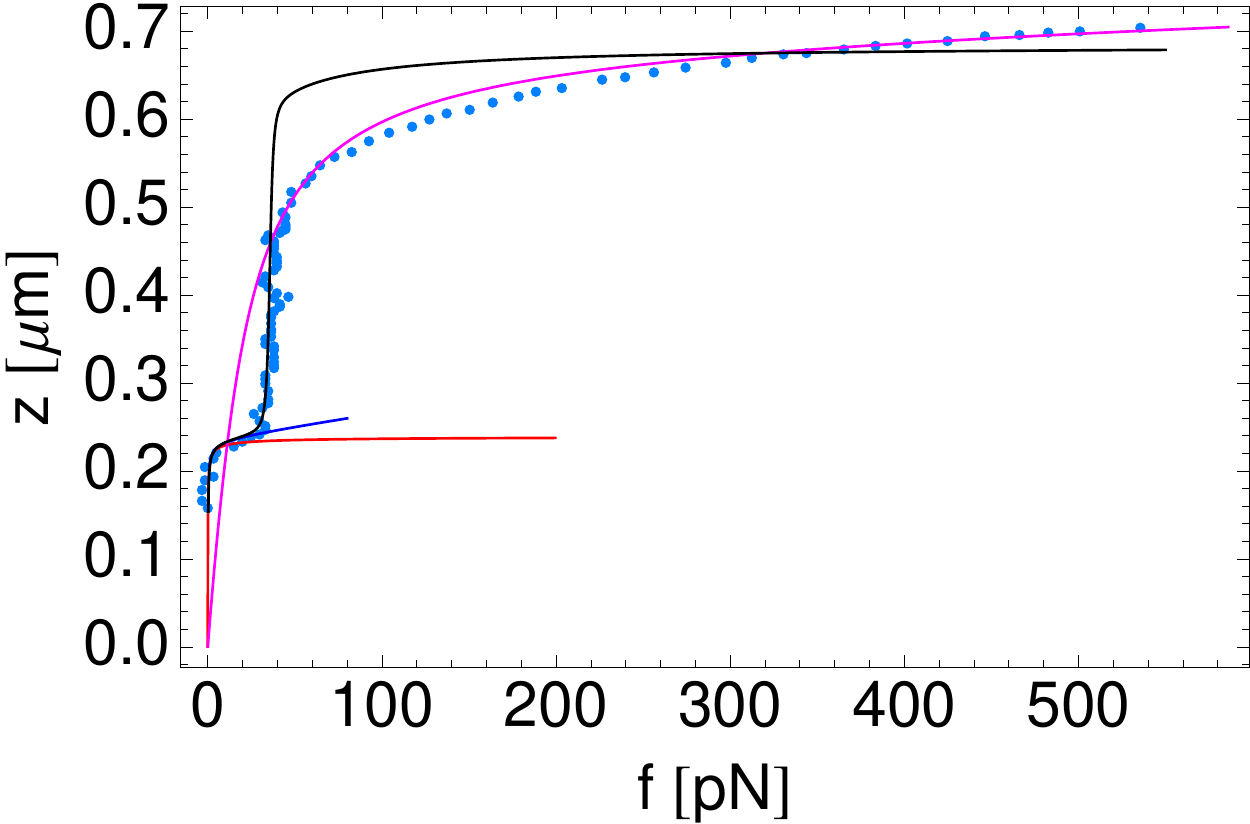}
\caption{Extension vs. force for a poly(dA-dT). Data (symbols) are  taken from Rief \textit{et al.}~\cite{rief}.  The solid curves correspond to the discrete Discrete Worm like chain interpolation, \eq{interp1}, for B-DNA without stretching (red), and with stretching (blue) and \eq{interp} ssDNA (pink) with a small bond size. The black curve correspond to \eq{extbond}, where linear stretching is included. Parameters values: $Na_B=0.24 \;\mu$m, $\kappa_B=147$, $\gamma=2.85$, $\kappa_{ss}=\kappa_{Bss}=0.75$, $b=0.285$, $\mu=5.2$, $J=1.5$, $E_B=800$.}
\label{polyAT}
\end{figure}

In Fig.~\ref{polyAT} is displayed an attempt to fit the transition observed for poly(dA-dT). Following Rief \textit{et al.}~\cite{rief}, we assume a transition from B-DNA to ssDNA. Hence the part of the curve after the transition is fitted by assuming that only a \textit{single} strand remains attached to the cantilever, the second free strand being splitted off. One thus finds a good fit (within experimental error bars) by keeping the same parameter values as for Fig.~\ref{figssDNA} but with a smaller $\kappa_{ss}=0.75$ (instead of 1.5 in Fig.~\ref{figssDNA}). This difference might be due to the different base-pair sequence. Two important remarks can be done. First, the ratio $\gamma=a_{ss}/a_{B}$ is around 2.85, which is significantly larger than the geometrical expected value of 2.1. It can be attributable either to the use of \eq{ext_homo} with a small Kuhn length, or to the fact that the $z=0$ reference was not well set in the experiment. Another explanation could be that the spontaneous curvature of the AT sequence~\cite{crothers} decreases the effective bp length in the B state. 

Second and more importantly, we are not able to properly fit the transition, especially the part of the curve which is after the transition using \eq{extbond} (black solid curve). This is due to the fact that, at the transition, the number of degrees of freedom increases since for ssDNA the effective bond length is $a=0.285\; a_{ss}$. Taking into account this fact would require drastically a different model. Hence we conclude that the  B to ssDNA transition cannot be explained by such a mesoscopic model where the monomeric unit remains unchanged through the transition. Writing a model where the number of monomers $N$ is force-dependent remains challenging.

\section{Conclusion}

This work intends to clarify some issues related to the mesoscopic modeling of DNA molecules subject to an external force, which can be applied, for example, by an optical tweezer or an AFM tip. We have focused on both ssDNA and dsDNA molecules because it has been suggested that dsDNA can denaturate under load, thus leading to two unpaired, single strands. 

We have first addressed the modeling of ssDNA under load and provide an analytical formula, \eq{interp}, that allows to obtain a very good fit on a wide range of forces, from 0 to 1~nN. Our main conclusion is that this molecule cannot be accurately modeled by a polymer, the monomer of which is a nucleobase. This finding is remarkable: whereas mesoscopic DNA models using the nucleobase as an elementary building block are usually relevant at small forces, the strong force regime requires the use of smaller, sub-nucleobase monomers. Under strong load, the constitutive chemical elements of a nucleobase play a significant role because they do not constitute a perfectly rigid entity. This is particularly important if one wishes to model the B to ss or S to ss transition: writing a mesoscopic model where the nature of the monomers and their number varies continuously with an external parameter (here the applied force) is challenging and, to our knowledge, has never been undertaken.

As far as dsDNA is concerned, we have proposed a new generalization of the extensible, discrete Marko-Siggia formula~\cite{Rosa,lipowsky} to a two-state model, which describes successfully force-extension transitions of semiflexible polymers. 
During their first transition, $\lambda$-phage or poly(dG-dC) DNAs remain in a duplex form, that we have called the ``S form'' as others, where the helix is unwound and successive bases are unstacked. \eq{extbond}, enables us to fit accurately experimental B to S transitions and its validity is corroborated by an exact transfer matrix approach (which is computationally more complex). These fits, done on different data sets of $\lambda$-DNA and poly(dG-dC) (Figs.~\ref{figrief} and \ref{lambdaSN}), yield consistently the same parameter values for the S-DNA bending rigidity $\kappa_S\simeq4$ and monomer length $a_S/a_B\simeq1.7-1.9$, and the Ising parameters, $\mu=4.5$ and $J=1.7$ for poly(dG-dC), and $\mu\simeq4$, $J=2$ for $\lambda$-DNA. The latter are consistent with the small but non-negligible influence of the base sequence.

In contrast, this model is not able to fit the transition observed for poly(dA-dT) (see Fig.~\ref{polyAT}). Our conclusion is that poly(dA-dT) is subject to a peeling transition where dsDNA is denaturated, thus confirming previous analysis~\cite{rief}. In this ss form, a single strand seems to remain under load. While the B to S transition can be accurately modeled because the basic unit (a base) of the model remains the same in the B and S forms, this is not the case for a B to ss transition. Note that for the S to ss transition observed for poly(dG-dC) at 320~pN, entropic effects are negligible and our model still yields an acceptable fit (see Fig.~\ref{polyGC2dtrans}(b), using the same parameters for ssDNA as in Fig.~\ref{figssDNA}) but not for $\lambda$-DNA. It must be emphasized that to conclude these points it was central to be able to fit ssDNA stretching curves on the nanoNewton range.

Put together, these results suggest that when increasing the strength of the base-pairing between both strands (at a given salt concentration), the nature of the transition changes. At low base-pairing strength, e.g. for poly(dA-dT), unpairing occurs at sufficient low forces so that both unpairing and unstacking transitions are simultaneous~\cite{Netz10}. At high enough base-pairing strength, in $\lambda$-phage or poly(dA-dT), unstacking occurs first, leading to the S form, and it is followed at stronger forces by unpairing, leading to the ss form. The effect of base sequence is much smaller for the B to S transition, maybe because it is essentially the base stacking which is modified during the transition.

In a very recent work, Zhang et al.~\cite{zhang} observed, using magnetic tweezers, that B-DNA has two different structural transitions around 60--70~pN, selected by the temperature or the salt concentration.
They differ thermodynamically by the sign of $\partial f_c/\partial T$ at the transition, slightly positive for the nonhysteretic (probably B to S) transition and positive for the hysteretic (B to ss) one. Our approach yields a negative slope for the B to S transition in the accessible temperature window, as seen in Fig.~\ref{polyGC2dtrans}(a), which seems contradictory. However, in our mesoscopic model, we did not consider solvent or counterions entropy which are implicitly included in our parameter $\mu$. Taken $\mu$ as a function of temperature might allow us to reconcile these two approaches. This work is in progress.
Our result, \eq{extbond}, may thus be useful to study, in systematic experiments, the role of the variation of salt concentrations~\cite{rouzina2} and base-sequence on stretching transitions. Since our study aims to bridge the gap between force-extension curves and thermal denaturation profiles, one thus could benefit from the huge quantity of works done in DNA melting~\cite{wartben,gotoh,polscher}.

Finally, we use equilibrium models for describing the transitions and do not consider effects of loading rates~\cite{rief,cocco,whitelam}, and hysteresis~\cite{fu1,fu2,zhang}. Note however that it has been shown that rehybridisation~\cite{ferrantini} of one strand or the closure of one denaturation bubble are very long processes of several $\mu$s depending on the DNA length, and one can expect that such equilibrium approaches remain valid at moderate loading rates.

\bigskip

\appendix
\section{How to fit ssDNA experimental force-extension curves}

We recall the equation derived in the text for the force-extension curve of ssDNA, $f(z)$:
\be
\frac{a f}{\kB T}= \frac{z}{L} (1+U_{\rm nl}(f)) \left(3\frac{1-u(\kappa)}{1+u(\kappa)}-\frac1{\sqrt{1+4\kappa^2}}\right)+\sqrt{\frac1{[1-z(1+U_{\rm nl}(f))/L]^2}+4\kappa^2}-\sqrt{1+4\kappa^2}
\label{interp_app}
\ee
where the ssDNA contour length $L$, the bending rigidity modulus $\kappa$ (in $\kB T$ units), and the ``effective'' monomer size $a$ (which can differ from the nucleobase length $a_{ss}\simeq 0.7$~nm) are the three unknown parameters. The Langevin function is $u(x)=\coth x -1/x$ and the non-linear stretching polynomial is $U_{\rm nl} (f)=1.172777\; f  - 3.731836\;f^2 +  4.118249\; f^3$ where $f$ must be in units of 10~nN.

Since \eq{interp_app} is highly non-linear, a convenient and extremely accurate simplification is to use $f$ defined by:
\be
\frac{a f_{\rm DMS} }{\kB T}= \frac{z}{L} \left(3\frac{1-u(\kappa)}{1+u(\kappa)}-\frac1{\sqrt{1+4\kappa^2}}\right)+\sqrt{\frac1{(1-z/L)^2}+4\kappa^2}-\sqrt{1+4\kappa^2}
\label{interp1_app}
\ee
in the argument of $U_{\rm nl}$ in \eq{interp_app}. Computing $f(z)$ is then very simple using, for instance, a spreadsheet.

\section{How to fit B to S-DNA transition in experimental force-extension curves}

The fitting procedure is the following: (i)~We fit the low force regime, $0\leq f\leq20$~pN, using \eq{interp1} (identical to \eq{interp1_app} above) by fixing the known B-DNA bending modulus $\kappa_B=147$~\cite{prl07,pre08}. This amounts to fitting the scale-parameter for the $z$ axis, the contour length of B-DNA $Na_B$ (this step is done here only for Fig.~\ref{figrief}(a) since the data have already been rescaled in the $z$-scale for the other data sets). (ii)~The stretching modulus $\tilde E_B$ is then determined by fitting the whole data before the transition ($0\leq f\leq60$~pN). (iii)~The ratio $\gamma=a_S/a_B$ and the bending modulus of the S form, $\kappa_S$, are fixed by fitting only the data after the transition using \eq{interp1_app}. (iv)~Finally the transition is fitted using \eq{extbond} which is:
\bea
\frac{z}{a_B N} = \left(1+\frac{F}{\tilde E_B}-\frac1{2\alpha_B}\right)\varphi_B+\gamma\left(1-\frac1{2\alpha_S}\right)\varphi_S +\frac{\lan\sigma_i\sigma_{i+1}\ran-1}4\left(\frac1{2\alpha_B}\frac{\kappa_B-\kappa_{BS}}{\kappa_{BS}+F/2+\alpha_B}+\frac\gamma{2\alpha_S}\frac{\kappa_S-\kappa_{BS}}{\kappa_{BS}+\gamma F/2+\alpha_S}\right)
\label{extbond_app}
\eea
where
\bea
\alpha_{B,S} &=& \sqrt{\kappa_{B,S}F +\left(\frac{F}2\right)^2}\\
\varphi_{B,S} &=& \frac12\left( 1\pm \frac{\sinh \mu_0}{\sqrt{\sinh^2\mu_0+e^{-4J_0}}}\right)\\
1-\lan\sigma_i\sigma_{i+1}\ran &=& \frac{e^{-4J_0}}{\sqrt{\sinh^2L_0+e^{-4J_0}}(\cosh L_0+\sqrt{\sinh^2L_0+e^{-4J_0}})}
\eea
and the effective Ising parameters are
\bea
\mu_0 &=& \mu-\ln\gamma+F\frac{1-\gamma}2+\frac12\ln\left(\frac{\kappa_S+F\gamma/2+\alpha_S}{\kappa_B+F/2+\alpha_B}\right)\\
J_0 &=& J+\frac14\ln\left[\frac{(\kappa_{SB}+F/2+\alpha_B)(\kappa_{SB}+F\gamma/2+\alpha_S)}{(\kappa_{B}+F/2+\alpha_B)(\kappa_{S}+F\gamma/2+\alpha_S)}\right] 
\eea
Since $Na_B$, $\gamma$, $\kappa_B$, $\kappa_S$ and $\tilde E_B$ are known thanks to steps (i)-(iii), this last step yields the values of the parameters $\mu$ and $J$, thus fixing the position (defined by $\mu_0=0$) and the width of the transition respectively. We have checked that choosing $\kappa_S=\kappa_{BS}$ does not change  significantly the results.

\end{document}